  \documentclass[preprint2]{aastex}

\def\lsim{\mathrel{\rlap{\lower 4pt \hbox{\hskip 1pt $\sim$}}\raise 1pt \hbox
        {$<$}}}
\def\gsim{\mathrel{\rlap{\lower 4pt \hbox{\hskip 1pt $\sim$}}\raise 1pt \hbox
        {$>$}}}

\shorttitle{Synthesized Mass of $^{56}$Ni}

\shortauthors{Umeda \& Nomoto}

\begin{document}
\title{How much $^{56}$Ni can be produced in Core-Collapse Supernovae? :
Evolution and Explosions of 30 - 100 $M_\odot$ Stars}

\author{Hideyuki Umeda$^{1}$ and Ken'ichi Nomoto$^{1, 2}$}

\affil{
$^{1}$Department of Astronomy, School of Science, University of Tokyo, Hongo,
Tokyo 113-0033, Japan}
\affil{
$^{2}$Institute for the Physics and Mathematics of the Universe, 
University of Tokyo, Kashiwa, Chiba 277-8582, Japan}
\email{umeda@astron.s.u-tokyo.ac.jp; nomoto@astron.s.u-tokyo.ac.jp}

\affil{\rm{ accepted for
publication in the Astrophysical Journal}}

\begin{abstract}

 Motivated by the discovery of extremely bright supernovae
SNe1999as and 2006gy, we have investigated how much 
$^{56}$Ni mass can be synthesized in core-collapse massive
supernovae (SNe). We calculate the evolution of several
very massive stars with initial masses $M \leq 100 M_\odot$
from the main-sequence to the beginning of
the Fe-core collapse and
simulate their explosions and nucleosynthesis. 
In order to avoid complications associated with strong mass-loss,
we only consider metal-poor stars with initial metallicity
$Z= Z_\odot/200$. However, our results are applicable to
higher metallicity models with similar C+O core masses.
The C+O core mass for the 100$M_\odot$ model is $M_{\rm CO}=
42.6M_\odot$ and this is the heaviest model in the literature
for which Fe-core collapse SN is explored.
The synthesized $^{56}$Ni mass increases 
with the increasing explosion energy and progenitor mass.
For the explosion energy of $E_{51} \equiv E/10^{51}$erg =30, for example, the 
$^{56}$Ni masses of $M(^{56}$Ni) =  2.2, 2.3, 5.0, and 6.6 $M_\odot$
can be produced for the progenitors with initial masses of 30, 50, 80
and 100 $M_\odot$ (or C+O core masses $M_{\rm CO}$= 11.4, 19.3, 34.0 and 42.6 $M_\odot$),
respectively. We find that producing $M(^{56}$Ni) $\sim 4M_\odot$ 
as seen in SN1999as is possible for $M_{\rm CO} \gsim$  34 $M_\odot$
and $E_{51} \gsim 20$. Producing $M(^{56}$Ni) $\sim 13M_\odot$ as suggested 
for SN2006gy requires
a too large explosion energy for $M_{\rm CO} \lsim 43M_\odot$,
but it may be possible with a reasonable explosion energy
if $M_{\rm CO} \gsim 60M_\odot$. 
\end{abstract}

\keywords{supernovae: general --- supernovae: individual (SN 1999as, SN 2006gy) 
--- nuclear reactions, nucleosynthesis, abundances}

\section{Introduction}

 Massive stars with initial masses of $M \sim 10 - 130M_\odot$ form
iron cores at the end of their evolution, and the collapse of
the iron cores triggers supernova (SN) explosions. During the explosion
explosive Si-burning produces radioactive isotope $^{56}$Ni.
The released energy by its decay to $^{56}$Fe via $^{56}$Co becomes
the dominant energy source to power the optical light of
most of SNe.
The ejected mass of the $^{56}$Ni is typically $M(^{56}$Ni) =$0.07 - 0.15 M_\odot$
for canonical core-collapse SNe, such as SN1987A and SN1993J 
(e.g., Nomoto et al. 2004 for a review).

 The explosion mechanism of core-collapse SNe
is still uncertain. At present most popular mechanism
is the delayed explosion model. In this model the gravitational 
energy released by the collapse is
first converted into neutrinos inside a proto-neutron star, then
the neutrinos heat up the matter outside 
to form an energetic shockwave. This mechanism may work in relatively
less massive SNe of $M \lsim 20 M_\odot$ to produce
the normal explosion energy of $\sim 10^{51}$ ergs.

 It has been recognized
that there are significant numbers of
core-collapse SNe, such as SN1998bw, 2002ap, 2003dh, 2003lw,
that explode with much larger explosion energies than
canonical SNe. These energetic SNe are called 
``hypernovae'', whose observational indication
is the very broad line spectral feature 
and relatively large ejected mass of $^{56}$Ni
(e.g., Galama et al. 1998; Iwamoto et al. 1998).
Theoretical modeling 
suggests that these SNe are more massive than 
$\sim 20 M_\odot$, and the compact remnant mass is large enough to 
exceed the maximum neutron-star mass (e.g., Nomoto et al. 2004).
 In modeling hypernovae we have found that the synthesized $M(^{56}$Ni) 
increases with explosion energies in the core-collapse
SNe models with a given progenitor mass. However, there may be
an upper limit for the $M(^{56}$Ni), 
because the reasonable range of the explosion energy is 
upper-limited and also larger explosion energy yields a larger
$^4$He to $^{56}$Ni ratio in the complete
Si-burning region (e.g., Nakamura et al. 2001; see also Figures 5 \& 6  below). 

 An interesting question is how much
$^{56}$Ni production is theoretically possible 
for core collapse SNe. One motivation to study this question is 
the observability of high-redshift SNe, that are
explosions of massive stars in the early universe.
In the early universe which has still very
low metallicity, it is expected that the initial mass function
could be top heavy or have a
peak around $\sim 100 - 200 M_\odot$ 
(e.g., Nakamura \& Umemura 1999; Abel et al. 2002).  
The stars more massive than $140
M_\odot$ become the pair-instability SNe (PISNe). 
However, we have shown in our previous
papers (Umeda \& Nomoto 2002, 2003, 2005; UN02, UN03, UN05 hereafter) 
that the nucleosynthesis patterns of very metal-poor stars
do not fit to those of PISNe (also, Heger \& Woosley 2002). 
Instead many of the stars correspond to hypernovae or
energetic Fe-core collapse SNe.
From precise comparisons, $M(^{56}$Ni) is estimated to be
typically $0.1 - 0.4 M_\odot$ (Tominaga et al. 2007b; UN05) 
and comparable to a typical present day hypernovae such as SN1998bw. 
Such hypernovae can be modeled by the explosion of $25-50M_\odot$ stars with
explosion energy $\sim 10-40\times 10^{51}$ ergs. 

 In the present days,
however, there are very interesting peculiar SNe, 
SN1999as (Knop, R., et al. 1999) and 
SN2006gy (Foley et al. 2006; Ofek et al. 2007; Smith et al. 2007).
These SNe are extremely bright, and the observational and theoretical preliminary 
results suggest that 
$M(^{56}$Ni) $\sim 4 M_\odot$ for SN1999as (Hatano et al. 2001)
and $M(^{56}$Ni) $\gsim 13 M_\odot$ for SN2006gy (Smith et al. 2007; Nomoto et al. 2007).

 Although these estimates are preliminary,
these $M(^{56}$Ni) are much more than those in previously known SNe. 
In Table 1, we summarize some properties of the presently
known hypernova and their candidates; i.e., SN type, estimated progenitor
mass, C+O core mass and the ejected $M(^{56}$Ni) obtained by spectral and
light curve fitting.

 As suggested by Smith et al. (2007),
we have tempted to regard those extremely bright SNe as PISNe.
However, the PISN models are difficult to reproduce the
broad light curve of SN2006gy (Nomoto et al. 2007).
In order to explain their light curves, much more explosion energy
or much less ejecta mass is required than the 
PISNe models currently available (UN02; Heger \& Woosley 2002). This 
provides us a motivation to study how much
$^{56}$Ni production is possible in the core-collapse SNe because 
hypernova models predict much narrower light curve than
PISNe models. 
 
 The synthesized $^{56}$Ni mass increases with the progenitor mass.
In the literature the most massive progenitor model available is the
$M=70M_\odot (M_{\rm He}=32M_\odot, M_{\rm CO}=28M_\odot)$ model
in Nomoto et al. (1997). Using this progenitor model we never obtain
$M(^{56}$Ni) $> 6M_\odot$ for the reasonable range of the explosion
energies. 

 Heger et al. (2003) discussed
how massive single stars end their life for almost entire mass and 
metallicity ranges. The models relevant to us is $M < 140M_\odot$ 
metal poor
stars and these are included in their discussion. However, no detailed
explanation for the progenitor models are given in the paper. They 
implicitly assumed that metal-poor $M \sim 50 - 30M_\odot$ stars,
i.e., massive C+O core, collapse directly to black holes and eject
almost no $^{56}$Ni. This may be true for the perfectly non-rotaing stars.
However, the explosion of rotating black hole forming
stars is uncertain and they may even explode as hypernovae
with large amount of $^{56}$Ni.

 In this paper, we thus calculate several very massive 
metal-poor stellar models up to 100 M$_\odot$ from the main-sequence 
to the beginning of the Fe core collapse.
Unlike Heger et al. (2003), we explore the possibility that these
stars explode energetically and using these models we investigate the $^{56}$Ni 
mass production as a function of the stellar mass and explosion energy.
Such very massive stars experience severe mass-loss when the 
stellar metallicities are large. Since the amount of mass-loss
is very uncertain, we only consider such metal-poor models as
$Z=10^{-4} =Z_\odot/200$, for which the mass loss rate and thus
the uncertainty in the stellar mass
is relatively small. We note that the way of explosion, including
the amount of the synthesized $^{56}$Ni, is basically independent
of the existence of He and H-envelopes. Therefore, most our results
are applicable to higher metallicity SN models with the
same C+O core progenitor masses.

 The explosion mechanism of hypernovae is yet unknown.
However, the inferred large remnant mass suggests that a black hole is
formed and thus the explosion mechanism is
different from the delayed neutrino heating
model. One possible mechanism is the 
formation of the central black hole \& accretion-disk system, and the
extraction of the energy from that; this is similar to the 
``micro-quasar'' (Paczy\'nski 1998),
``collapsar'' models (MacFadyen \& Woosley 1999; 
MacFadyen, Woosley \& Heger 2001) and the jet-like explosion
(Maeda \& Nomoto 2003a,b). If this is the case, we need to perform
2D or 3D calculations of the explosion. 
For simplicity, however, we 
perform only 1D calculations assuming spherical symmetry.
The spherical explosion model gives the upper limits to 
$M(^{56}$Ni) because in the jet-like
explosion, for example, $^{56}$Ni may be synthesized only
along the jet directions.

 In section 2, we describe our model and the calculation method.
In section 3 we show the results of the presupernova stellar
evolutions and the synthesized $M(^{56}$Ni) during the SN
explosions. Section 4 gives conclusions and discussions.

\section{Model and Calculations}

 We calculate stellar evolution from the main-sequence to the
pre- Fe-core-collapse using the same method as adopted in UN02, i.e.,
a Henyey type stellar evolution code which is fully 
coupled to a nuclear reaction network to calculate energy
generation and nucleosynthesis. Some description of
the code is also given in Umeda et al. (2000) and UN05.

 We adopt the solar chemical composition by Anders \& Grevesse (1989),
the solar He abundance $Y_\odot =0.27753$, and metallicity
$Z_\odot =0.02$. We assume that the initial helium abundance depends
on metallicity $Z$ as $Y(Z)= Y_p +(\Delta Y/\Delta Z)Z$, where
primordial He abundance
$Y_p=0.247$ and $ \Delta Y/\Delta Z$=1.5265. In this paper, we
only deal with $Z=10^{-4}=Z_\odot/200$ models. We adopt the
empirical mass loss rate (de Jager et al. 1988) scaled with the metallicity as 
$(Z/0.02)^{0.5}$ (Kudritzki et al. 1989). 

 As has been
addressed by many authors (e.g., Weaver \& Woosley 1993;
Nomoto \& Hashimoto 1988; Imbriani et al. 2001)
one of the most important quantities that affect the pre-SN stellar structure 
and nucleosynthesis is the carbon abundance after He-burning. 
This quantity sensitively depends on the treatment for the core convection
during He burning and the reaction rate
$^{12}$C($\alpha,\gamma)^{16}$O. Therefore, we take the reaction
rate as a parameter. According to our previous paper, we
multiply the $^{12}$C($\alpha,\gamma)^{16}$O rate of
Caughlan \& Fowler (1988; CF88 hereafter) by a constant factor 
as shown in Table 2.  

 As for the treatment of convection, we assume Schwartzshild criterion
for convective instability and diffusive treatment for the convective
mixing. The time variation of the abundance $X_i$ of a chemcal element $i$,
is governed by the equation
\begin {equation}
 \left( \frac{\partial X_i}{\partial t} \right)_{M_r} = \left[ 
\left( \frac{\partial X_i}{\partial t} \right)_{M_r} \right]_{\rm nuc} 
 - \frac{\partial }{\partial M_r} \left(D \frac{\partial X_i}{\partial M_r} \right),
\end {equation}
where [...]nuc represents the change due to nuclear reactions,
the second term of the right hand side represents the effect
of convective mixing. In this code we first solve the [...]nuc
part simultaneously with hydrodynamics, and after convergence
calculate convective mixing by the second term of Eq.(1)
For the diffusion coefficient $D$ of the 
convective mixing we adopt the form derived for semi-convection
by Spruit (1992):
\begin{equation}
 D = \rm{max}(0, \nabla_{\rm rad}-\nabla_{\rm ad}) f_k K_t 
 (4 \pi r^2 \rho)^2 \left( \frac{4}{\beta}-3 \right) / \nabla_\mu,
\end{equation}
where $\beta$ is the ratio of gas pressure to total pressure $\beta = p_{\rm gas}/p$,
$\nabla_\mu$ is the logarithmic gradient of mean molecular weight,
\begin{equation}
 \nabla_\mu= \frac{d \rm{ln} \mu} {d \rm{ln} p}, 
\end{equation}
$K_t$ is the thermal diffusivity defined by 
\begin{equation}
 K_t = \frac{4 a c T^3}{3 \kappa \rho^2 c_p}, 
\end{equation}
and $f_k$ is the squre root of the 
ratio of ``solute'' to thermal diffusivity. If molecular diffusivity is 
used to estimate $f_k$, the value becomes very small as discussed in 
Spruit (1992). Because we are not certain what are the physical process
occurring in a star, we treat this quantity as a parameter (Saio \&
Nomoto 1998, unpublished; Iwamoto \& Saio 1999; Umeda et al. 1999).

 Strictly speaking the above coefficient (Eq.(2)) cannot be used for the
standard convective region where $\mu$ gradient is nearly zero. In such a
region we replace $\nabla_\mu$ by a small finite constant (say $10^{-6}$)
such that the coeeficient $D$ takes nearly the same order as the mixing 
length theory. This formalism is useful when we study the effects of
semi-convection and convective over-shootings though we do not consider
such effects in this work. The same coefficient is adopted for our previous
works since 1999; Iwamoto \& Saio (1999) and their later works (e.g., 
Iwamoto et al. 2004).

For most of the models
in this paper we assume relatively slow 
convective mixing $(f_k=0.1$; Umeda et al. 2000). Larger $f_k$
leads to stronger mixing of nuclear fuel, thus resulting in stronger 
convective shell burning. This effect leads to a smaller mass 
core. However, larger $f_k$ leads to less carbon fraction after 
He-burning, which instead leads to a larger core. Therefore, it is 
not simple to predict the final outcome as a function of the parameter
$f_k$.

  Once pre-SN progenitor models are obtained, we simulate their SN
explosions with a 1D piecewise parabolic method (PPM) code by putting
explosion energy into some mesh points just below the mass cut. 
Here we do not specify the 
explosion mechanism and assume spherical symmetry for simplicity
because the hypernova mechanism is highly uncertain. The detailed
nucleosynthesis during the explosion is calculated by postprocessing
as described in UN02. 

\section{Results}

\subsection{Pre-Supernova Evolution}
 
 In Table 2, we summarize the parameters and 
some properties of our pre-SN progenitor models.
The first three columns are the initial mass, convection
parameter $f_k$ and the factor multiplied to the CF88
$^{12}$C($\alpha,\gamma)^{16}$O rate. The next six columns are
the pre-SN progenitor final-mass with mass-loss ($M_f$),
He core mass, C+O core mass, C/O mass fraction after He burning,
Fe core mass and the binding energy 
$E_{\rm bin}$ of the layer above the Fe-core, 
respectively. Here the He, C+O and Fe core masses are 
defined by the mass coordinate where the fractions of hydrogen and helium 
are X(H)$<10^{-3}$,  X(He)$<10^{-3}$
and electron mole fraction $Y_e < 0.49$ in the progenitor models,
respectively. 

 The total amount of mass-loss is larger for more massive stars.
However, with this low metallicity the total lost mass is not so large 
even for $M = 100M_\odot$ and a large fraction of
the hydrogen envelope still remains. Therefore, 
such a star becomes SN II unless it is in a binary system
with its envelope being stripped off by binary interaction.

 In the models shown in this paper, the Fe core mass is 
larger for more massive stars. This is usually the case but
not always, because the Fe core mass depends
on the C/O ratio after He-burning. For a larger C/O ratio,
convective C-burning is stronger, thus forming a smaller Fe-core
(e.g., Nomoto \& Hashimoto 1988).
The C/O ratio is generally smaller for more massive stars,  
because the $^{12}$C($\alpha,\gamma)^{16}$O
reaction is more efficient than the 3$\alpha$ reactions
at higher temperatures and lower densities. However, the C/O
ratio also strongly depends on the convective mixing at the
end of He-burning.

 The evolutions of $M \lsim 100 M_\odot$ stars are all similar
in the sense that all these stars form Fe-cores to initiate
Fe-core collapse. Nevertheless, we find a difference 
in the evolution of the stars more massive than $80M_\odot$.
These stars experience  ``pulsations'' at the end of 
Si-burning (see also Heger et al. 2003).

 In Figure 1 we show the evolution of the central density and
temperature for the 30 and 90$M_\odot$ models. It is well known
that a more massive star has a larger specific entropy at the center. 
Thus the temperature of the more massive star is higher for the same
density.  Along with this fact we find that the central temperature and density
evolution is very different between the 30 and 90$M_\odot$ models
during the ``Si''-burning stage, which takes place around the 
temperature $T_ 9$ = $2.5-4$. Here ``Si'' includes not only 
$^{28}$Si but also all the elements 
heavier than Si and lighter than Fe. The central temperature
and density of the 90$M_\odot$ model
oscillate several times. This is because in such
a massive star radiation pressure is so dominant that the adiabatic
index of the equation of state is close to $\Gamma \sim 4/3$, 
and thus the inner core of the stars easily
expands with the input of the nuclear energy released by ``Si''-burning.
Once it expands, the temperature drops suddenly, central
``Si''-burning stops and the stellar core turns into contraction.
Since only a small amount of ``Si'' is burnt in each cycle, these
pulsations occur many times.
Note that as shown in Figure 1 the density - temperature
track is very close to (but outside of)
the electron-positron
pair-instability region'' where $\Gamma < 4/3$
(e.g., Barkat et al. 1967). 

 A similar type of oscillations has been reported for metal-poor
stars of $ 100M_\odot \gsim M \gsim 140M_\odot $ 
in Heger et al. (2003) as 
"Pulsational PISNe". They discussed that these stars encounter
pair instability, leading to 
violent pulsations, but not complete disruptions.
Each pulse can have up to several $10^{51}$ ergs,
and only outer layers of the star are expelled. In our model for
the $80-100M_\odot$ stars we have not found such mass-loss
probably because the mass-range is lower and also we have attached
H-envelope that absorbs and relieves the pulsations.

 We have found from the study of 80$ - 100 M_\odot$ stars that
the number of the oscillations depends on the convective parameter
$f_k$: larger $f_k$ increases the number of the oscillation.
This is because for a larger $f_k$
fresh Si is more efficiently mixed into the central region, 
which increase the lifetime of this stage. 
The amplitude of the temperature and density
variation is larger for more massive stars, which suggests that more
and more drastic oscillations could occur for stars
of $M > 100M_\odot$.
 During this stage, the specific entropy at the center
is reduced (or density is increased)
and the Fe-core, which is defined as a region with $Y_e \leq 0.49$, grows.
However, the pulsations does not much affect on $M(^{56}$Ni).

\subsection{Pre-supernova Density Structure}

 As described in the next subsection, 
the amount of produced $^{56}$Ni is closely related to the
density structure, more specifically the enclosed mass ($M_{\rm r}$) -
radius ($r$) relation, of the pre-SN progenitors, where
$M_{\rm r}$ is the mass contained within a sphere of radius $r$.
Figure 2 shows this $M_r$ - $r$ relation for our progenitor models. 
For reference, in Figure 3 and 4 we show distributions of density $\rho$ and 
electron mole fraction $Y_e$, respectively. 
We see that the $M_r$ - $r$ curve is 
steeper for more massive stars, which means more massive
stars are more centrally concentrated. 
In some cases, however, less massive stars have a steeper
$M_r$ - $r$ relation if the C/O ratio after the
He burning is smaller, which leads to a larger core mass.
 
\subsection{Explosive Nucleosynthesis of $^{56}$Ni}

 The products of explosive burning are largely determined by
the maximum temperature behind the shock, $T_s$. Material
which experiences $T_{s,9} \equiv T_s/10^9$ K $> 5$ undergoes complete
Si-burning, forming predominantly $^{56}$Ni. The region
after the shock passage is radiation dominant, so that
we can estimate
the radius of the sphere in which $^{56}$Ni is dominantly 
produced as 
\begin{equation}
 R_{\rm Ni} \sim 3700 {E^*}_{51}^{1/3} ~ { \rm km},
\end{equation}
which is derived from $E^*=4\pi R_{\rm Ni}^3 a T_s^4/3$ with
$T_{s,9}=5$ (e.g., Thielemann et al. 1996).
Note that the deposited energy $E^*$ is approximately equal to the
sum of the explosion energy $E_{\rm exp}$
and the binding energy ($E_{\rm bin} >0)$
of the progenitor, i.e., $E^* \simeq E_{\rm exp} + E_{\rm bin}$.

We show in Figure 2 the radii which correspond to 
$ E^*_{51}=1, 10, 50$ and 100, where
the solid circles on the $M_{\rm r}$ - $r$ curves 
indicates the outer edge of the core, i.e., 
the masses of Fe-cores. 
 This figure suggests that the $^{56}$Ni synthesis 
is larger for larger $ E^*$ and for
the models with steeper $M_{\rm r}$ - $r$ curves, i.e., more massive stars. 
The mass between the outer edge of the
Fe-core and $r=R_{\rm Ni}$ gives the crude upper-limit to 
$M(^{56}$Ni) in the ejecta. 
This is an upper-limit because
the mass-cut, that divides the compact remnant and ejecta,
can be far above the Fe-core, and also 
the mass fraction of $^{56}$Ni in the complete Si-burning
region can be significantly smaller than 1.0 especially for the very
energetic explosion (see Figure 5 \& 6 and discussion below). 

 In order to obtain the synthesized $M(^{56}$Ni) more 
precisely, we perform explosion simulation using the 1D
PPM code and detailed postprocess nucleosynthesis 
calculations (see e.g., UN05).
Figures 5 \& 6 show examples of the results obtained in this way.
These are the post-explosion abundance
distributions in the 50$M_\odot$ and 100 $M_\odot$ models for the explosion
energy of $E_{51}=30$ (left) and $E_{51}=100$ (right), respectively. 
As shown in these figure, for a more energetic explosion,
$\alpha-$rich freezout is stronger so that
the mass fraction ratio of $^{4}$He/$^{56}$Ni increases
in the complete Si-burning region. 

 The results are summarized in Table 3. Here, the models 
A-F are the same as in Table 2. For each model, the results
with various explosion energies are shown.
 The upper limit of the $^{56}$Ni mass is shown as 
$^{56}$Ni$_{\rm up}$ in the last column. This is the upper
limit, because the mass-cut is assumed to be equal to the
Fe-core mass given in Table 2. In reality, though it 
depends on the model assumptions\footnote
{If the explosion is strictly spherical and the location of
the energy deposition is uniquely determined, we can uniquely
determine the mass-cut as a function of $E_{\rm exp}$.
However, in an aspherical model or in a jet-like explosion
with energy injection by continuous jets 
(e.g., Tominaga et al. 2007a), the "effective
mass-cut" may not simply the function of $E_{\rm exp}$.}
the mass-cut is likely larger than the Fe-core mass,
then the ejected $^{56}$Ni mass could be smaller. $^{56}$Ni$_{\rm up}$
is also summarized in Figure 7.

 In Table 3 we show the compact remnant mass $M_{\rm rem}$
for reference. $M_{\rm rem}$ corresponds to the mass-cut
when 0.07 $M_\odot$ of $^{56}$Ni is ejected. 
If $M(^{56}$Ni) is larger, therefore, the remnant mass would be
smaller than $M_{\rm rem}$ shown in Table 3.

 In the 7 and 8th columns in Table 3,
we also show $M(^{56}$Ni) (almost exactly the same as the
ejected Fe mass for these low metalicity models) when the
mass ratios of Mg/Fe and O/Fe are the solar ratios, 
i.e., when [Mg/Fe] = 0  and [O/Fe] = 0, respectively. 
We show these quantities, because in the observations of
metal-poor halo stars, most stars have the abundance
of [Mg/Fe] $\geq $ 0  and  [O/Fe] $\geq $ 0. This means that 
if SNe yield [Mg/Fe] $ < $ 0 or [O/Fe] $< $ 0, or
producing larger $M(^{56}$Ni) than in the 7 and 8th columns,
such SNe should not be dominant in the early Galaxy. 
We note that if we constrain the ejecta abundances
as [O/Fe] $\geq 0$, 
$^{56}$Ni ejection of more than $3M_\odot$ is not possible for
all the models considered in this paper.

 We, however, should make a caution
that the O and Mg yields from massive stars depend on
the treatment of convection (e.g., $f_k$) and the adopted 
$^{12}$C($\alpha,\gamma)^{16}$O reaction rate.
The model dependency of O yield would be small, at most 
a factor of 2, but that of Mg yield would be much larger 
and the yield can be a factor of 5 different at most. 
Generally the Mg yield increases when the
C/O ratio after the He-burning is larger because Mg is the
C-burning product.

 We find from Table 3 that if relatively less massive SNe 
($M \lsim 30 M_\odot$) produce $^{56}$Ni as much
as 2 $M_\odot$, the ejecta is too Fe-rich to
be compatible with metal-poor star abundance. On the other
hand, the ejection of $M(^{56}$Ni) $\sim 2M_\odot$ 
from such massive stars as $M \gsim 80 M_\odot$   
does not contradict with the abundance of metal-poor stars.

 We note that large $M(^{56}$Ni) ($\gsim 4 M_\odot$) models are
quite energetic, so the Zn/Fe ratios in the ejecta are
relatively large, exceeding
the solar ratio. This is a distinctive
feature from any PISNe models which do not produce large Zn/Fe
(UN02). In the energetic explosions Co/Fe
ratios are also relatively high. However, as shown in UN05,
the Co/Fe ratio is not much enhanced for $Y_e \lsim 0.5$
even for the energetic explosions. In the present models,
indeed, the Co/Fe ratios are less than the solar value.
More detailed study about nucleosynthesis and its
parameter dependences are given elsewhere.

\section{Conclusions and Discussions}

 Motivated by the discovery of unusually bright SNe 2006gy and
1999as, we have investigated how much $M(^{56}$Ni) 
can be synthesized by core-collapse massive SNe
with the initial mass of $M \leq 100 M_\odot$. 
Observed properties of SN1999as and 2006gy suggest that 
the ejected $^{56}$Ni masses are $M(^{56}$Ni)
$\sim 4 M_\odot$ (Hatano et al. 2001) and 13 $M_\odot$ 
(Nomoto et al. 2007).
Previously the only known SNe model
that can produce such a large $M(^{56}$Ni) is the PISNe which
are the explosion of $140 - 300 M_\odot$ stars. However,
the light curves of PISNe may be too broad to reproduce the
observed light curves of SNe1999as and 2006gy, suggesting that a 
less massive star explosion (Nomoto et al. 2007). 

 We, therefore, have calculated the evolution of several
very massive stars with the initial masses of $M \leq 100 M_\odot$
from the main-sequence to just before the Fe-core collapse.
In order to minimize the uncertainty in the mass-loss, we
have adopted low-metal ($Z=Z_\odot/200$) models throughout
this paper. However, most results, especially the estimate
of the ejected masses of $^{56}$Ni and $^{16}$O,
are applicable to higher metallicity models with similar
C+O core masses. 
Using these newly calculated progenitor models,
we simulated the SN explosion of such massive
stars and calculated the mass of the synthesized $^{56}$Ni,
and other elements.

 We find that the $M \gsim 80 M_\odot$ stars ``pulsate''
during the central Si-burning stages. The existence of this stage 
makes the lifetime of the stars longer and makes the Fe-core grow. 
However, the Fe-core growth during this stage
is not the main reason why these massive stars can produce
such large $M(^{56}$Ni). In this sense the existence of this 
stage is not critical for the production of the large 
$M(^{56}$Ni).

 We have shown that the synthesized $M(^{56}$Ni) mass increases 
with the increasing explosion energy, $E_{\rm exp}$, 
and the progenitor mass $M$.
Among them the effect of $M$ is more important
than the energy because more massive stars tend to have steeper
$M_{\rm r}$ - $r$ curves in the density structure
of the progenitors, so that $M(^{56}$Ni) increases steeply
with $M$ for the same $E_{\rm exp}$. Larger $E_{\rm exp}$ leads
to a larger $^{56}$Ni producing region according to the
Equation (1); however, this also leads to a smaller Ni/He
ratio in that region. As a result, a large
$E_{\rm exp}$ does not result in so large $M(^{56}$Ni).
For $E_{\rm 51}$ = 30,
$M(^{56}$Ni) = 2.2, 2.3, 5.0, and 6.6 $M_\odot$
can be produced
for $M=$ 30, 50, 80 and 100 $M_\odot$, respectively, or 
for C+O core masses of $M_{\rm CO}$= 11.4, 19.3, 34.0 
and 42.6 $M_\odot$, respectively.

 It is possible to produce $M(^{56}$Ni) $ \sim 4M_\odot$
as seen in SN1999as if $M_{\rm CO} \sim 34M_\odot$
and $E_{51} \gsim 20$. On the other hand,
producing $M(^{56}$Ni) $ \gsim 13M_\odot$ as seen in SN2006gy is challenging
for core collapse SN models if $M_{\rm CO} < 43M_\odot$. In the present study,
the most massive model has $M_{\rm CO}$= 42.6 $M_\odot$. For this model,
the explosion energy of $E_{51} \sim 200 $ is required to
produce $M(^{56}$Ni) $ \sim 13M_\odot$ but such $E_{51} $ may be
unrealistically large, although there is no observational constraints.

 Nevertheless, we should note that 
if we simply extrapolate our results
to more massive stars up to $M \sim 130M_\odot$, 
$M_{\rm CO}$ is as large as $60M_\odot$.
Such a model would undergo core-collapse and 
produce $M(^{56}$Ni) $ \sim 13M_\odot$ without
too large explosion energy. Currently we are constructing such an
extreme model, and will present the results in a forth coming paper. 
We are also studying very massive stars above 300$M_\odot$
for which much study has not been done yet. Those 
stars may be too heavy to explode as PISNe and 
form blackholes (e.g., Fryer et al. 2001). 
The collapse of these stars may or
may not accompany aspherical mass ejection and potentially eject
$\sim 10M_\odot$  $^{56}$Ni (Ohkubo et al. 2006), though the stellar
metallicities must be very low in order to leave 
sufficiently massive cores. 

\acknowledgments

 We would like to thank J. Deng and N. Tominaga for useful discussions.  
This work has been supported in part by the grant-in-Aid for Scientific Research
(18104003, 18540231) and the 21st Century COE Program
(QUEST) from the JSPS and MEXT of Japan.

\clearpage

\begin{deluxetable}{llccccc}
\tabletypesize{\scriptsize}
\tablecaption{Properties of hypernovae and their candidates. 
\label{tbl-1}}
\tablewidth{0pt}
\tablehead{
\colhead{Name} & \colhead{Type} & \colhead{$E_{51}$} & 
\colhead{$M_{\rm ini}/M_\odot$} & 
\colhead{$M_{\rm C+O}/M_\odot$} & 
\colhead{M($^{56}$Ni)/M$_\odot$} &
\colhead{Reference}
}  
\startdata 
SN1997ef & Ic & $\sim 20 $   & $\sim 35$   & 10.0 & 0.15 
         & Iwamoto et al. 2000 \cr 
SN1998bw & Ic & $\sim 40$    & $\sim 30$  & 13.8  & 0.5  
         & Iwamoto et al. 1998 \cr 
SN2002ap & Ic & $\sim 7$     & $\sim 21$     & 3.3 & $0.10$ & Mazzali et al. 2002 \cr 
SN2003dh & Ic & $\sim 30-50$ & $\sim 35-40$  & 4.5 & $0.35 $ 
         & Mazzali et al. 2003 \cr 
SN1999as & Ic & $\sim 20-50$ & $ \sim 60-80$   & $\sim 20-30$ & 4 & Hatano et al. 2001  \cr 
SN2006gy & Ib/c ?  & $ ? $ & ?   & ? & $\gsim 13$ & Nomoto et al. 2007  \cr 
\hline
SN1987A & II & 1  & $\sim 18-20$  & 3 & 0.07 &  Nomoto et al. 2004\cr 

\enddata
\tablecomments{Supernova name, type, expected explosion 
energies in $10^{51}$ erg, 
initial progenitor 
masses, C+O core masses, ejected $^{56}$Ni masses, and the 
references are shown. One canonical SN
(1987A) is also shown for comparison.}
\end{deluxetable}

\clearpage

\begin{deluxetable}{cccccccccc}
\tabletypesize{\scriptsize}
\tablecaption{Parameters and properties of the pre-SN models.
\label{tbl-2}}
\tablewidth{0pt}
\tablehead{ \colhead{Model}&
\colhead{$M$} & \colhead{$f_k$} & \colhead{CF88$\times$} & 
\colhead{$M_{\rm f}$} & \colhead{$M$(He)}  &
\colhead{$M$(CO)} & \colhead{C/O} &  
\colhead{$M$(Fe)} & \colhead{$E_{\rm bin}$}
}
\startdata 
 & 20 & 0.05 &1.4 & 19.94 & 6.76 & 4.39 & 0.19/0.80 & 1.51 &1.63 \\
 & 20 & 0.10 &1.7 & 19.93 & 6.45 & 5.29 & 0.16/0.83 & 1.55 &1.38 \\
A & 25 & 0.10 &1.0 & 24.87 & 8.35 & 6.33 & 0.29/0.70 & 1.52 &1.83 \\
B & 30 & 0.10 &1.0 & 29.61 & 13.1 & 11.4 & 0.30/0.68 & 1.86 &1.87 \\
C & 50 & 0.50 &1.0 & 48.55 & 21.8 & 19.3 & 0.16/0.79 & 2.21 &3.67 \\
D & 80 & 0.10 &1.0 & 75.81 & 36.4 & 34.0 & 0.20/0.74 & 2.75 &5.46 \\
E & 90 & 0.50 &1.3 & 83.53 & 37.4 & 36.5 & 0.087/0.82 & 3.15 &4.73 \\
F & 100 & 0.10 &1.0 & 83.14 & 46.5 & 42.6 & 0.23/0.72 & 3.22 &7.19 \\
\enddata
\tablecomments{Each column is the
initial mass ($M$), convection parameter ($f_k$),
the factor multiplied to the CF88 $^{12}$C($\alpha,\gamma)^{16}$O rate,
pre-SN progenitor mass with mass-loss ($M_f$), 
He core mass, C+O core mass, C/O mass fraction after He burning,
Fe-core mass and the binding energy
$E_{\rm bin}$. Here, all masses are shown in the units 
of $M_\odot$ and the energy is in $10^{51}$ erg.
These models are all for the
metallicity $Z=Z_\odot/200$.
}
\end{deluxetable}

\clearpage

\begin{deluxetable}{ccccccccc}
\tabletypesize{\scriptsize}
\tablecaption{The masses of Mg,  O and upper mass limits of the
$^{56}$Ni in the ejecta of core-collapse SNe  
as a function of progenitor mass and explosion energy. 
\label{tbl-3}}
\tablewidth{0pt}
\tablehead{ \colhead{Model}& \colhead{$M$}& 
\colhead{$E_{51}$} & \colhead{$M_{\rm rem}$}
& \colhead{Mg}& \colhead{O}& 
\colhead{$^{56}$Ni$_{\rm [Mg/Fe]=0}$}&
\colhead{$^{56}$Ni$_{\rm [O/Fe]=0}$} &
\colhead{$^{56}$Ni$_{\rm up}$} 
}
\startdata 
A & 25 & 1 & 2.11 & 0.14 & 3.50 & 0.34 & 0.46 &0.56 \\
B & 30 & 1 & 2.60 & 0.41 & 5.45 & 0.98 & 0.72 &0.69 \\
  &    & 10 & 3.57 & 0.41 & 4.68 & 0.98 & 0.62 &1.43 \\
  &    & 20 & 4.25 & 0.36 & 4.27 & 0.86 & 0.56 &1.83 \\
  &    & 30 & 4.77 & 0.31 & 3.86 & 0.75 & 0.51 &2.19 \\
  &    & 50 & 5.60 & 0.26 & 3.36 & 0.62 & 0.44 &2.52 \\
C & 50 & 1  & 2.46 & 0.79 & 12.52 & 1.90 & 1.65 &0.83 \\
  &    & 10 & 4.22 & 0.73 & 11.67 & 1.77 & 1.54 &1.48 \\
  &    & 30 & 5.44 & 0.79 & 10.73 & 1.91 & 1.42 &2.28 \\
  &    & 50 & 6.26 & 0.80 & 9.96 & 1.94 & 1.32 &2.74 \\
  &    & 70 & 6.90 & 0.80 & 9.49 & 1.92 & 1.25 &2.97 \\
  &    & 100 & 7.62 & 0.78 & 8.91 & 1.87 & 1.18 &3.19 \\
  &    & 150 & 8.62 & 0.76 & 8.16 & 1.82 & 1.08 &3.60 \\
D & 80 & 1 & 6.08 & 1.41 & 18.71 & 3.40 & 2.47 &2.14 \\
  &    & 10 & 8.00 & 1.35 & 17.45 & 3.25 & 2.30 &3.36 \\
  &    & 30 & 10.61 & 1.28 & 15.31 & 3.09 & 2.02 &4.99 \\
  &    & 50 & 11.89 & 1.20 & 14.33 & 2.89 & 1.89 &5.74 \\
  &    & 60 & 12.24 & 1.17 & 14.07 & 2.82 & 1.86 &6.06 \\
  &    & 100 & 15.04 & 0.88 & 12.09 & 2.11 & 1.60 &7.86 \\
E & 90 & 1  & 6.08 & 0.38 & 20.59 & 0.92 & 2.72 &1.85 \\
  &    & 10 & 7.82 & 0.36 & 19.69 & 0.86 & 2.60 &2.69 \\
  &    & 20 & 9.42 & 0.33 & 18.67 & 0.80 & 2.46 &3.51 \\
  &    & 30 & 10.75 & 0.31 & 17.84 & 0.74 & 2.36 &4.19 \\
  &    & 50 & 11.75 & 0.28 & 17.01 & 0.68 & 2.25 &4.85 \\
  &    & 70 & 13.07 & 0.26 & 16.00 & 0.62 & 2.11 &5.78 \\
 &    & 110 & 14.71 & 0.23 & 14.87 & 0.55 & 1.96 &6.97 \\
F & 100& 1  & 8.74  & 1.58 & 22.41 & 3.81 & 2.96 & 3.34\\
  &    & 30 & 14.22 & 1.34 & 19.27 & 3.34 & 2.54 & 6.64\\
  &    & 70 & 17.62 & 1.38 & 16.67 & 3.33 & 2.20 & 8.78\\
  &    & 100 & 19.81 & 1.28 & 15.31 & 3.08 & 2.02 & 10.17\\
  &    & 210 & 24.43 & 0.83 & 12.50 & 2.00 & 1.65 & 13.49\\

\enddata


\tablecomments{Models A-F are the same as in Table 2.
Columns 4 to 9 are the following masses in units of $M_\odot$:
(4) the remnant mass for the case of 0.07$M_\odot$ $^{56}$Ni 
ejection ($M_{\rm rem}/M_\odot$), (5, 6)
the total Mg and O masses in the ejecta, (7, 8)
the $^{56}$Ni masses when the ejecta has the solar abundance 
ratios for Mg/Fe and O/Fe ($^{56}$Ni$_{\rm [Mg/Fe]=0}$ and 
$^{56}$Ni$_{\rm [O/Fe]=0}$), respectively, 
and (9) the upper limit to the ejected $^{56}$Ni mass 
when the mass-cut is set just above the
Fe-core shown in Table 2 ($^{56}$Ni$_{\rm up}$)
as a function of progenitor mass $M$
and explosion energy $E_{51}\equiv E/10^{51}$ erg.}
\end{deluxetable}

\clearpage

\begin{figure}
\plotone{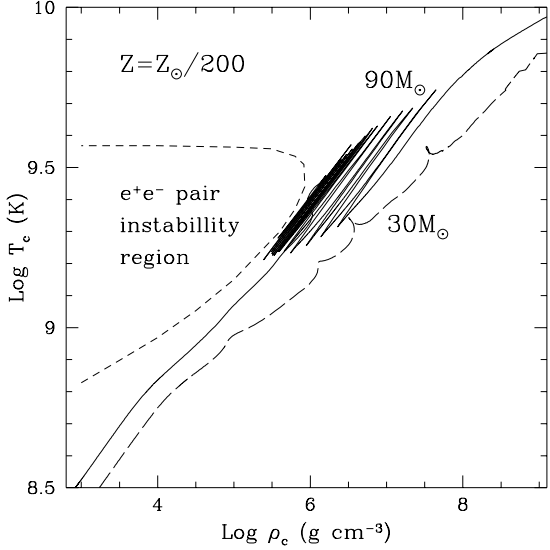}
\caption{Evolution of the central density and
temperature for the 30 and 90$M_\odot$ models.
}
\end{figure}

\begin{figure}
\plotone{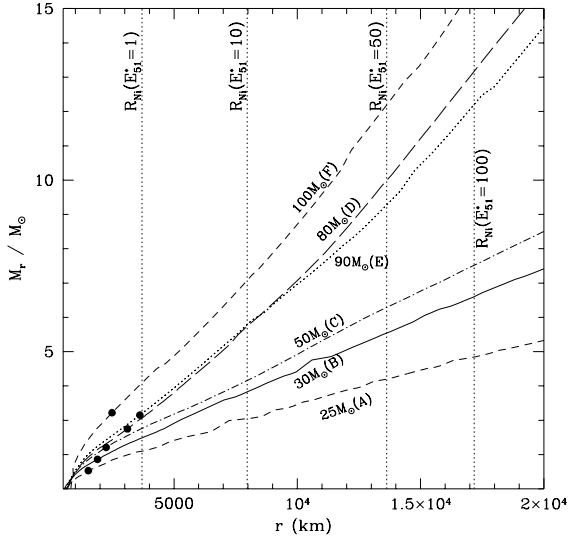}
\caption{The enclosed mass ($M_r$) -
radius relation of our pre-SN progenitors. The parameters
and some properties of the models A-F are shown in Table 2.
For each model the location of the top of the Fe-core is
shown by a filled circle. $R_{\rm Ni}$ is the upper radius
for the region where Ni$^{56}$ is dominantly produced
as a function of the deposited energy $E^*$
given by the Equation (5) in the text.
}
\end{figure}

\begin{figure}
\plotone{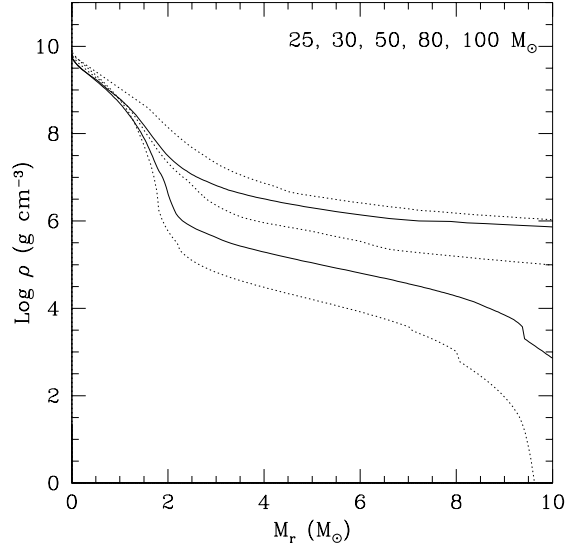}
\caption{The enclosed mass ($M_r$) -
density ($\rho$) relation of our pre-SN progenitors. 
The masses of the models are 25, 30, 50, 80, and
100 $M_\odot$, from bottom to top respectively. 
}
\end{figure}

\begin{figure}
\plotone{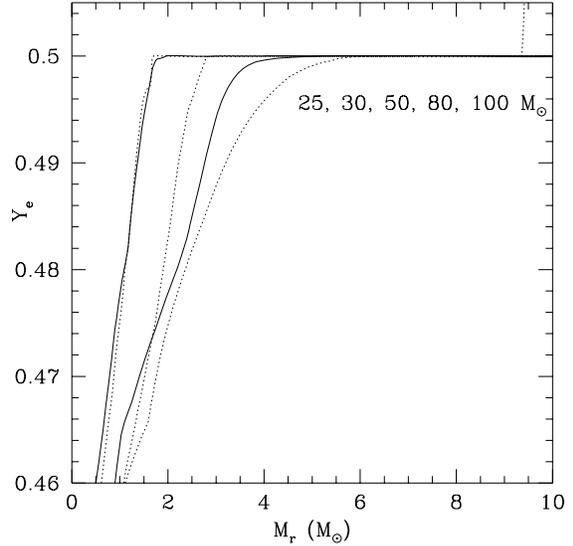}
\caption{The enclosed mass ($M_r$) -
electron mole fraction ($Y_e$) relation of our pre-SN progenitors. 
The masses of the models are 25, 30, 50, 80, and
100 $M_\odot$, from left to right respectively. Solid lines
represnt 30 and 80$M_\odot$ models.
}
\end{figure}

\begin{figure}

\plottwo{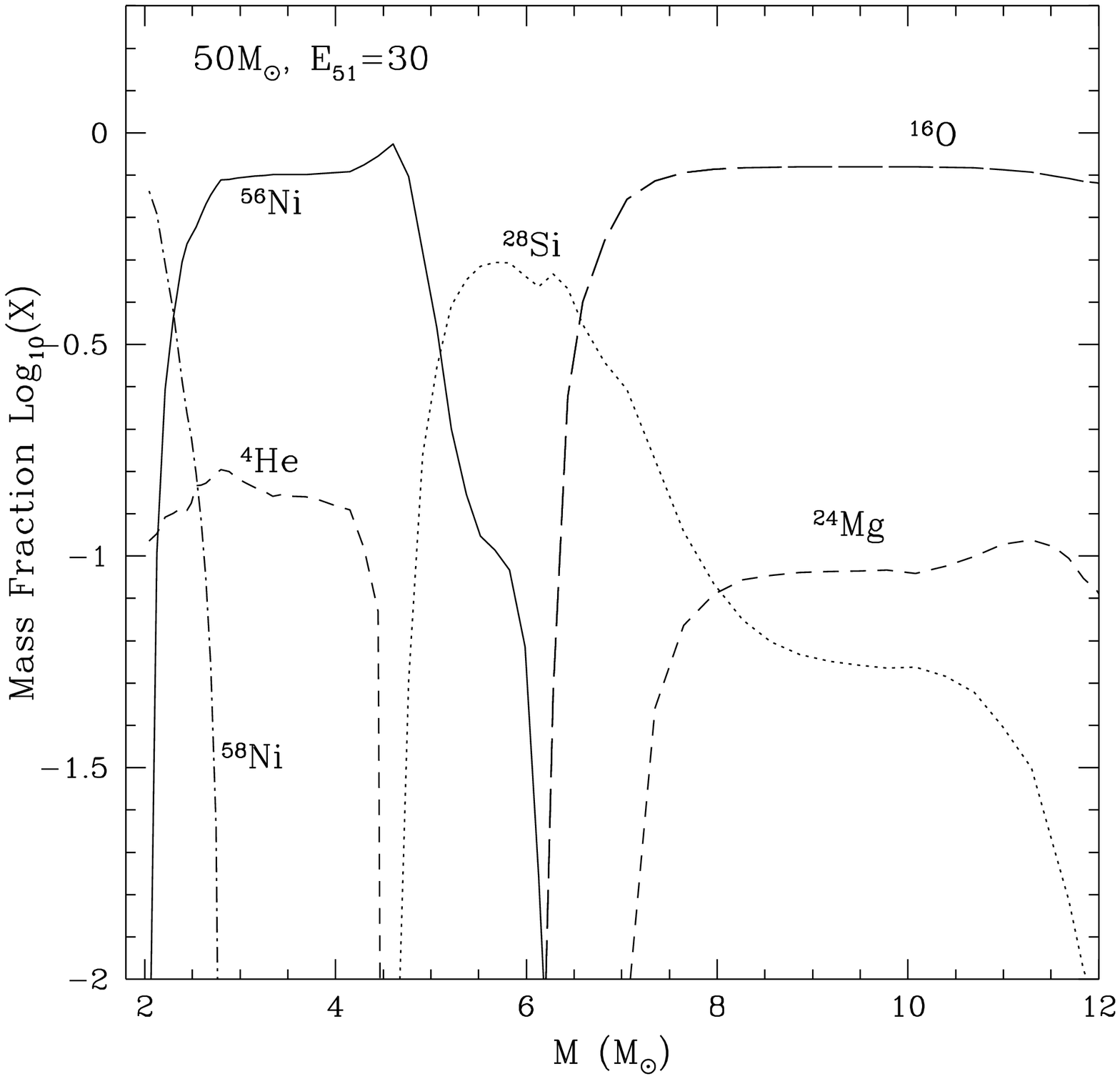}{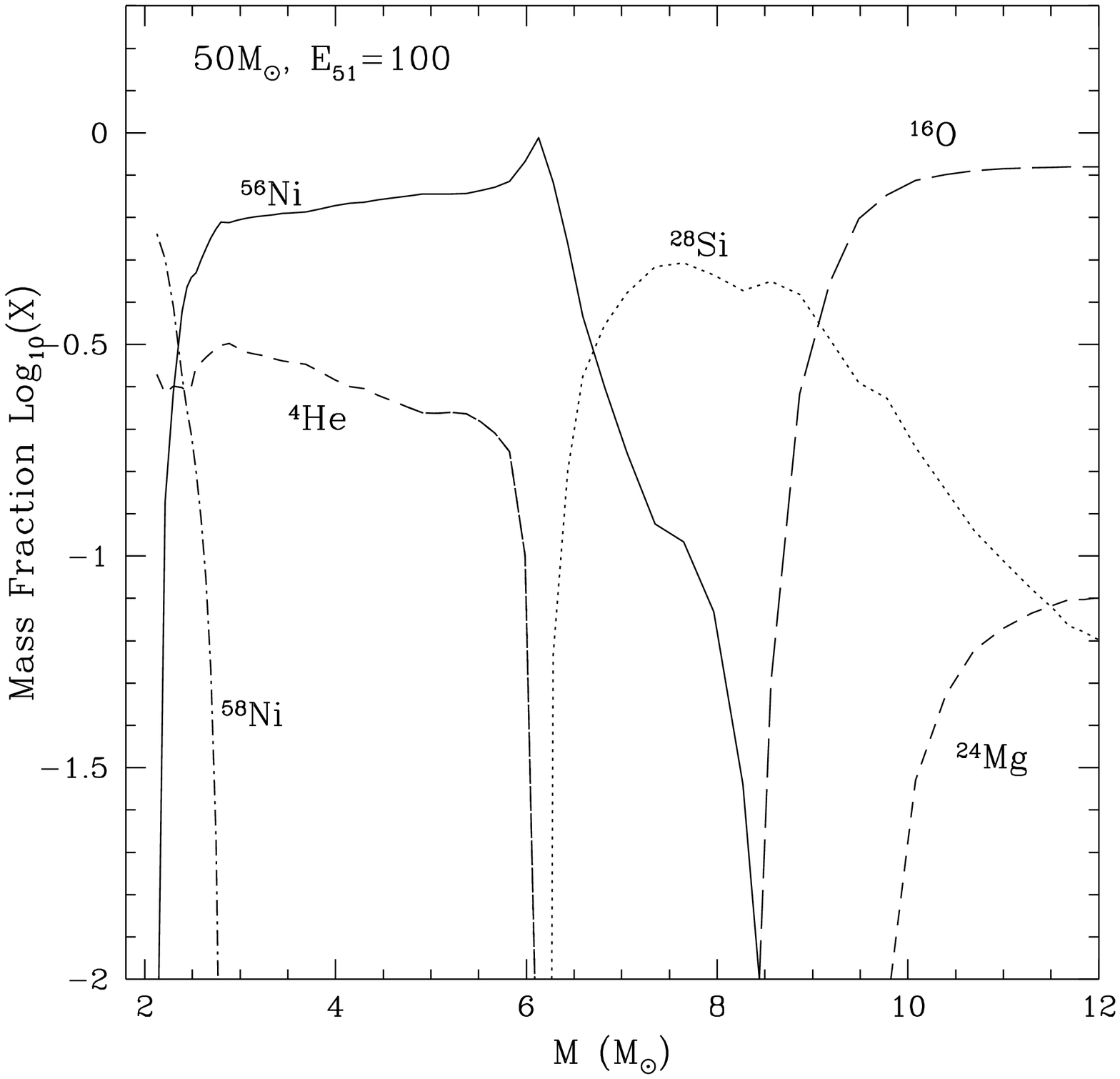}

\plottwo{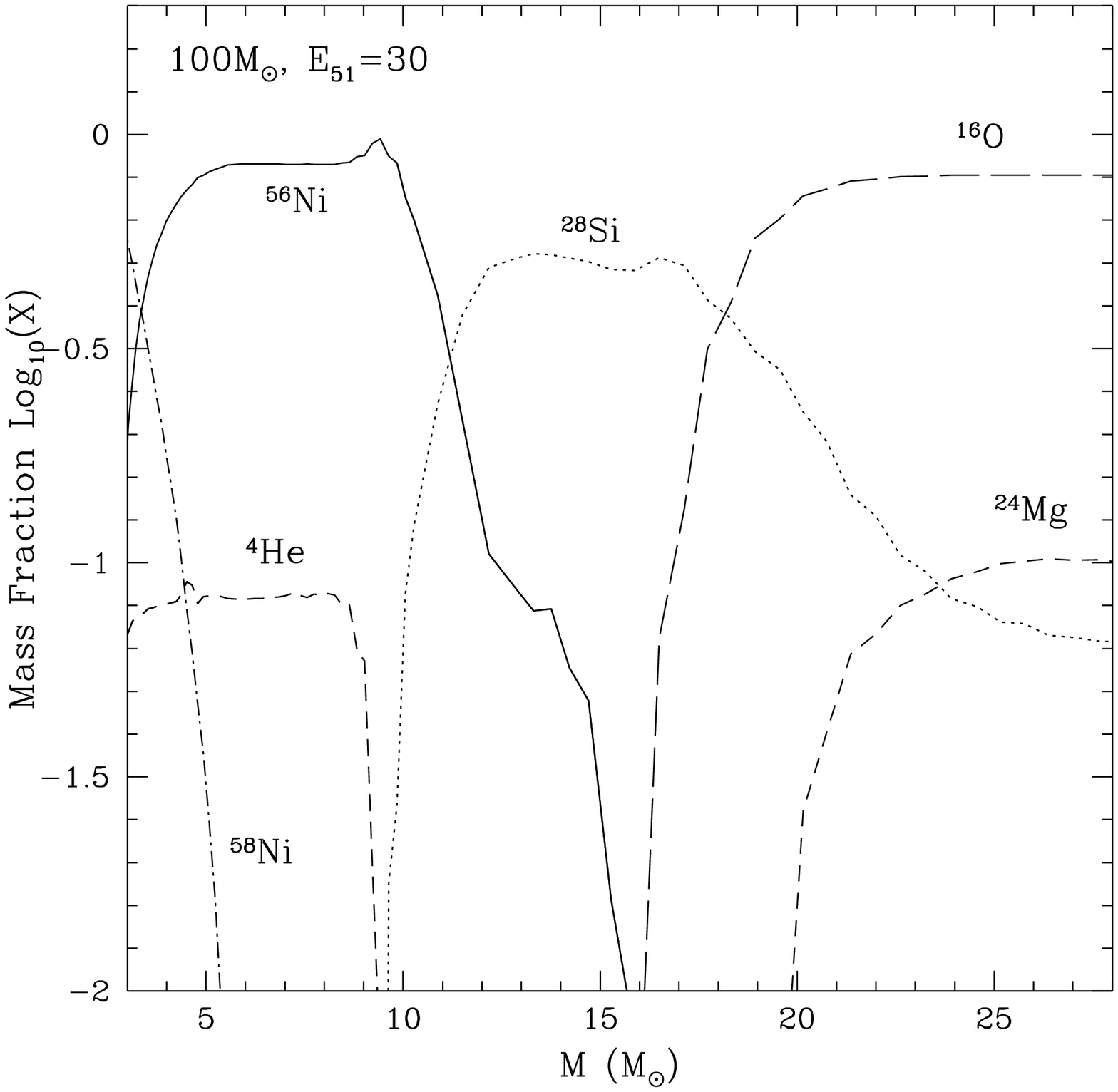}{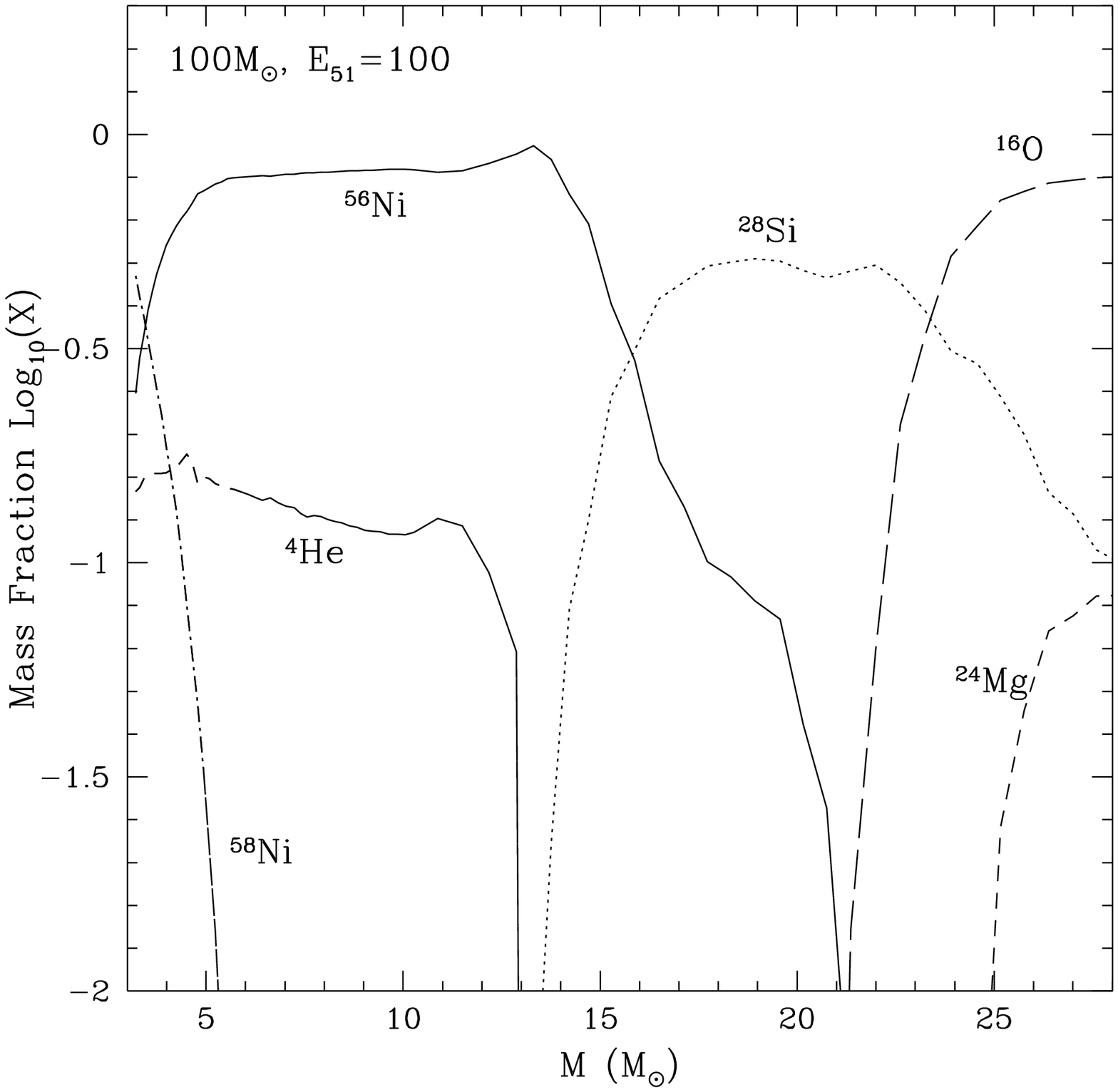}

\caption{The post-explosion abundance
distributions for the 50$M_\odot$ models (upper two panels)
and 100$M_\odot$ models (lower two panels) when the explosion
energy is $E_{51}=30$ (left) and $E_{51}=100$ (right.) 
}
\end{figure}

\begin{figure}
\plottwo{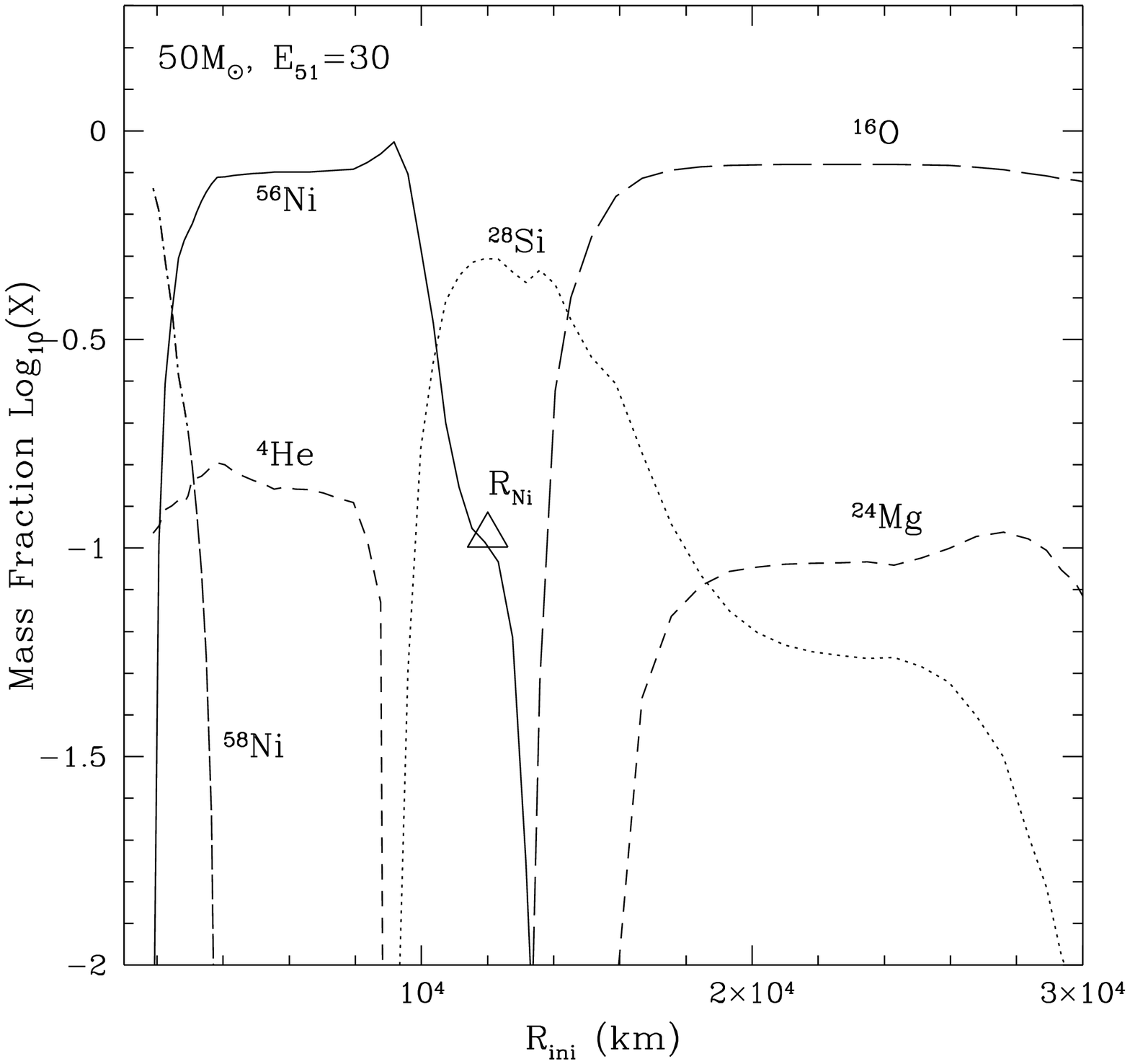}{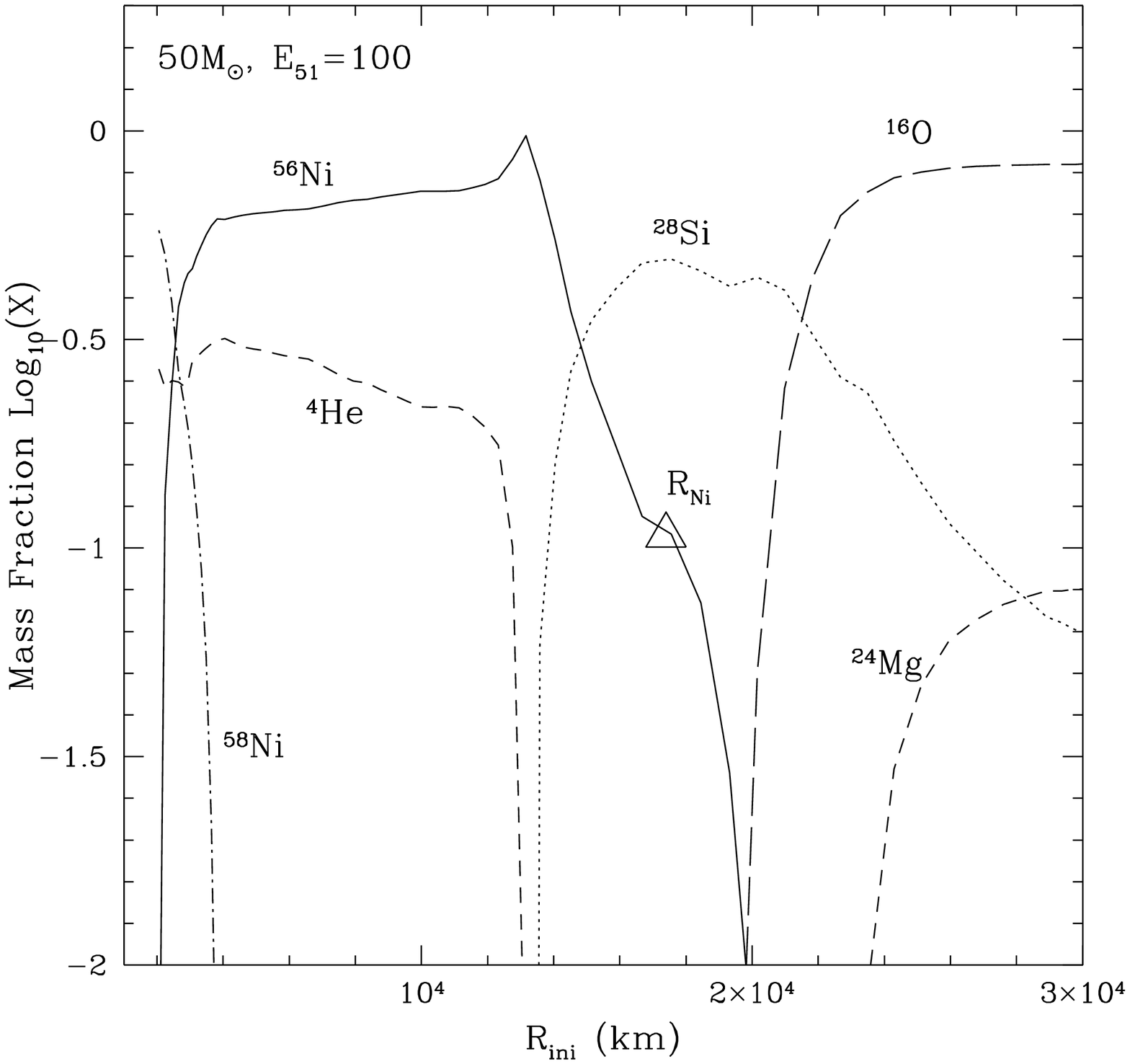}

\plottwo{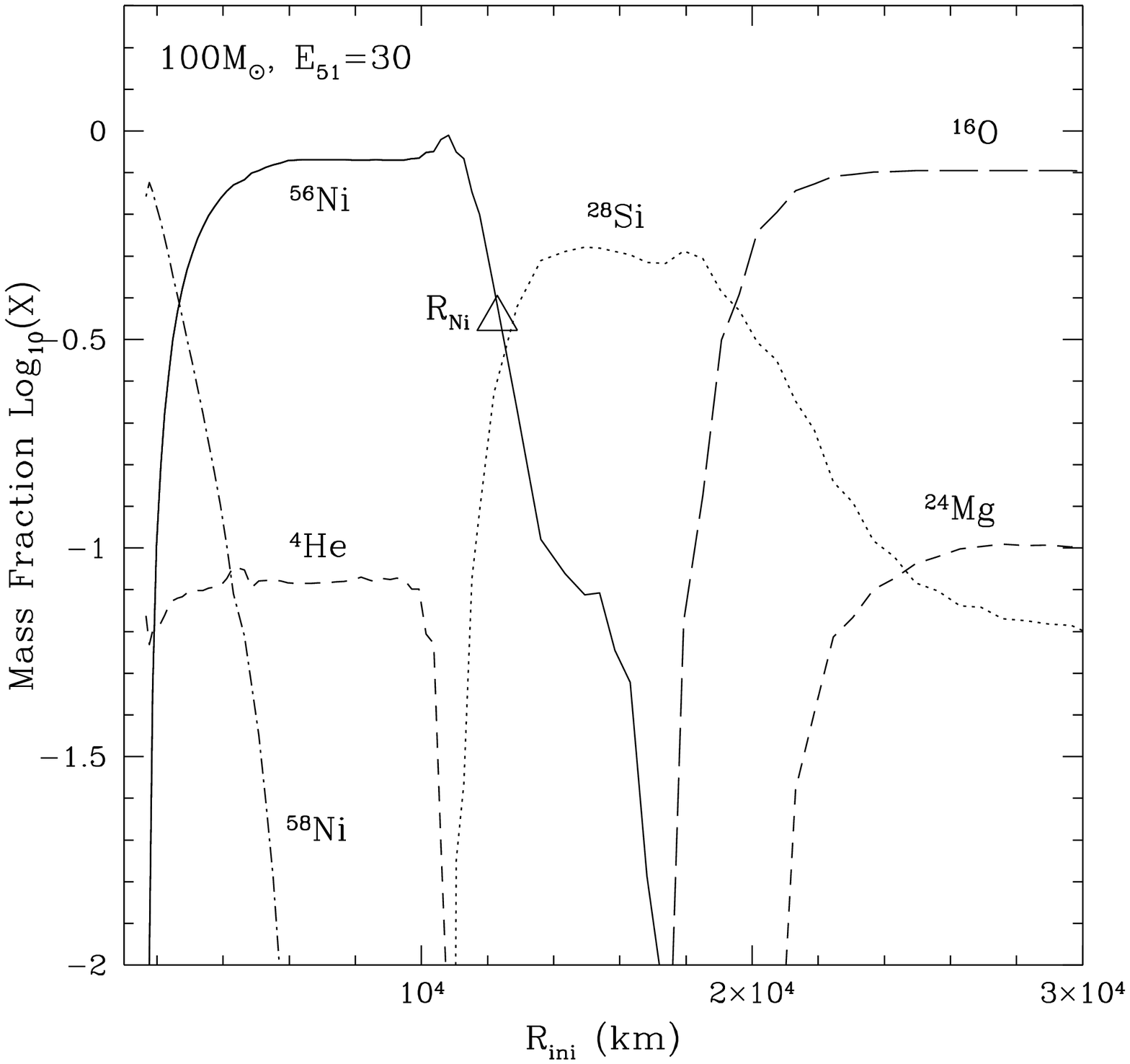}{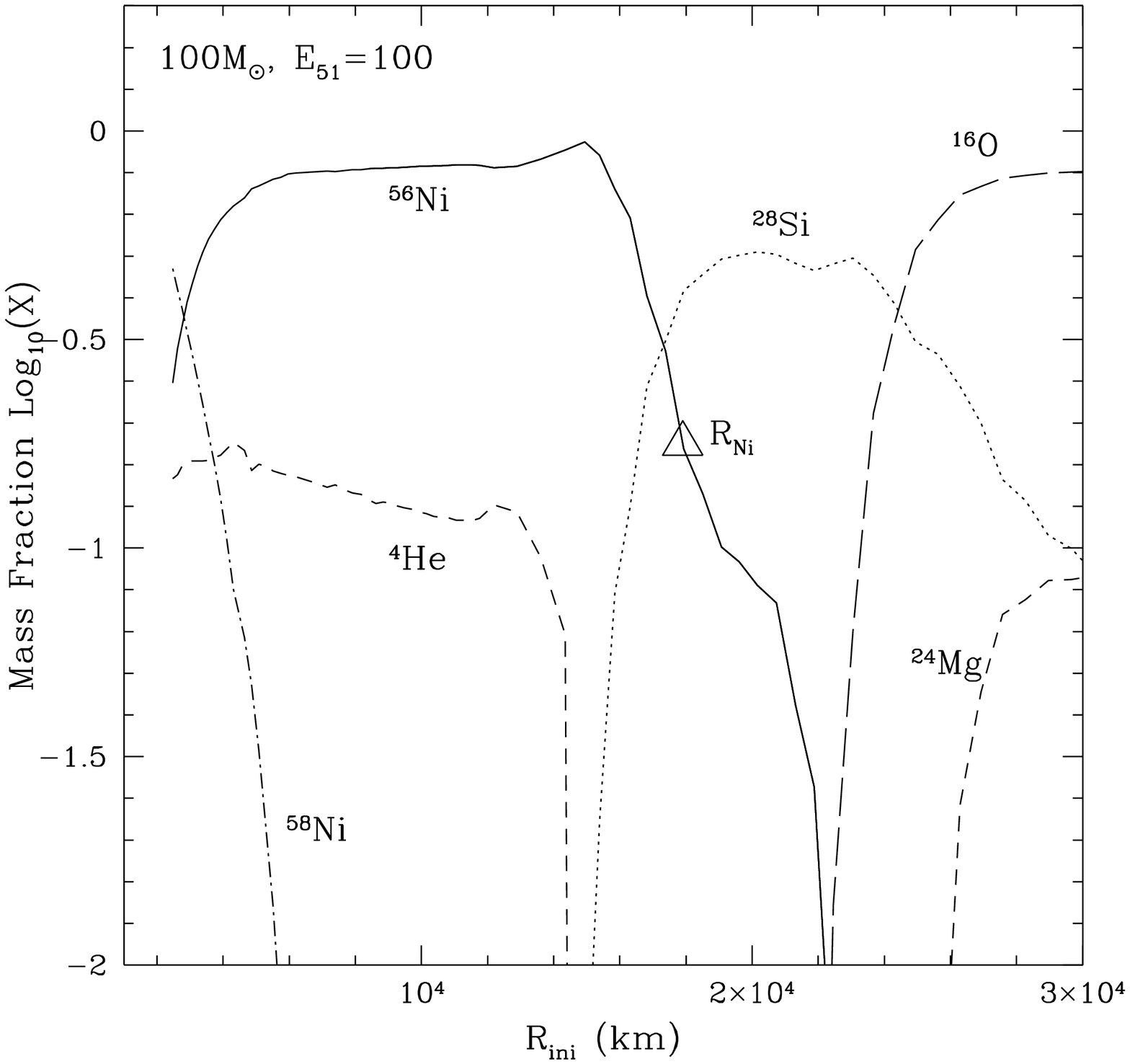}
\caption{Same as Figures 5 but the horizontal axis is replaced
by the radius of the progenitor, $R_{\rm ini}$ (km). The location 
of $R_{\rm Ni}$ calculated by the equation (5) are also shown by 
large triangles. Here, we note that $R_{\rm Ni}$ 
appears to be a little over-estimated because the temporal energy
absorption due to the matter dissociation is not taken into account 
in the equation (5).
}
\end{figure}

\begin{figure}
\plotone{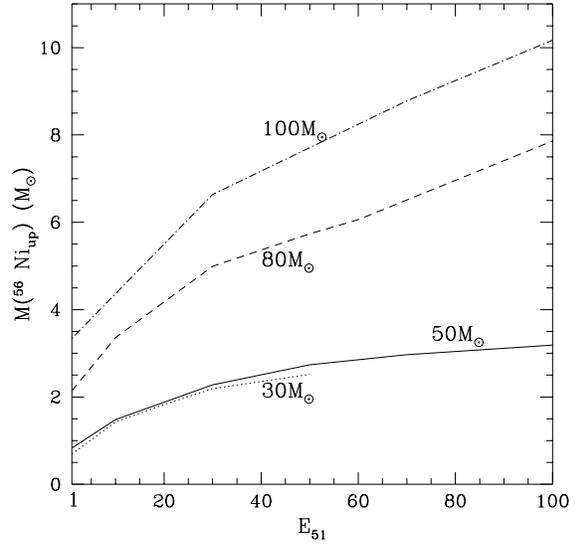}
\caption{The upper limit of $^{56}$Ni masses produced by core-collapse
SNe with various masses and explosion energies. $^{56}$Ni$_{\rm up}$ 
shown here are the ejected $^{56}$Ni mass when the mass-cut is
located at the top of the Fe-core. If mass-cut is larger the
ejected mass will be smaller.
}
\end{figure}

\end{document}